\documentclass[]{article}

\usepackage{graphicx}
\usepackage{authblk}
\usepackage{amsmath}
\usepackage[numbib]{tocbibind}
\begin{document}


\title{Generalized Einstein's and Brinkman's solutions for the effective viscosity of nanofluids} 



\author[1,2]{Y.O. Solyaev}
\author[1,2]{S.A. Lurie}
\author[1]{N.A. Semenov}

\affil[1]{Institute of Applied Mechanics of Russian Academy of Sciences, Moscow, Russia}
\affil[2]{Moscow Aviation Institute, Moscow, Russia}

\setcounter{Maxaffil}{0}
\renewcommand\Affilfont{\itshape\small}

\date{\today}



\maketitle 

\begin{abstract}
In this paper, we derive the closed form analytical solutions for the effective viscosity of the suspensions of solid spheres that take into account the size effects. This result is obtained using the solution for the effective shear modulus of particulate composites developed in the framework of the strain gradient elasticity theory. Assuming incompressibility of matrix and rigid behavior of particles and using a mathematical analogy between the theory of elasticity and the theory of viscous fluids we derive the generalized Einstein's formula for the effective viscosity. Generalized Brinkman's solution for the concentrated suspensions is derived then using differential method. Obtained solutions contain single additional length scale parameter, which can be related to the interactions between base liquid and solid particles in the suspensions. In the case of the large ratio the between diameter of particles and the length scale parameter, developed solutions reduce to the classical solutions, however for the small relative diameter of particles an increase of the effective viscosity is predicted.  It is shown that developed models agree well with known experimental data. Solutions for the fibrous suspensions are also derived and validated.
%
\end{abstract}

\section{Introduction}
\label{int}

Investigation of physical and mechanical properties of nanofluids attracts a lot of attention during last decades due to promising opportunities that it provides in heat transfer applications, in medicine, in solar energy, in MEMS/NEMS etc. \cite{devendiran2016review, taylor2013small}.

Theoretical description of the effective properties of nanofluids bases on the approaches developed in continuum mechanics, statistical mechanics, molecular dynamics \cite{Mahian2019, Meyer2016}. One of the first model have been developed in this area by Einstein \cite{einst1096} (see also \cite{landau1959}), whose formula for the dilute suspensions viscosity as well as its generalizations for the finite concentrations \cite{Brinkman1952,Batchelor1974} are widely used now for the characterisation of different colloidal systems. However, the particles size is neglected in these classical solutions, such that they cannot describe the known experimental data with nanofluids, which properties are significantly affected by the size effects \cite{Koca2018, Munyalo2018, Murshed2017}. Corresponding enriched models for the nanofluids have been developed recently accounting for Brownian motion \cite{Masoumi2009}, effects of agglomeration \cite{
krishna2015modeling,Ganguly2009}, slip velocity \cite{Jang2007}, the equivalent molecular diameter of base fluid \cite{Corcione2010}, and influence of particles capping layers \cite{Hosseini2010}.

In the present paper we propose the theoretical interpretation for the size dependent viscosity of nanofluids that bases on the idea that around the small particles the velocity gradients become rather high and the classical constitutive equations for such processes become inaccurate. Namely, the assumption that the viscous stresses depend only on the first spatial derivatives of velocity is no longer valid (this assumption is introduced in classical fluid mechanics only for the case of small velocity gradients, see e.g. Landau and Lifshitz\cite{landau1959}, p. 44). In general, this means that for the correct continuum description of such processes one should take into account that the rate of energy dissipation due to viscosity depends not only on the strain rate but also on the gradients of strain rate. Such type of the second gradient fluid mechanics for the small scale flows have been developed recently \cite{Fried2006, seppecher2000second} and widely investigated\cite{kim2009impact,fried2008continuum,dell2015postulations, abisset2015second}. 

In the present work, we do not solve directly the fluid mechanics problems, however, we use the analogy\cite{Batchelor1974} between mathematical formulation of the quasi-static elasticity theory and the theory of newtonian viscous liquids for the small Reynolds number (slow creeping flow), which can be also established in the case of gradient theories. Thus, we consider the well known Mindlin-Toupin strain gradient elasticity theory (SGET)\cite{Mindlin1964}, in which the strain energy depends on the strain and on the strain gradients. This theory is widely used today, e.g. for the analysis of the spatial dispersion of elastic waves in solids \cite{Askes2009,eremeyev2019comparison}, size dependent behavior of nanobeams\cite{Cordero2016,lurie2018revisiting}, effective properties of nanocomposites \cite{lurie2011eshelby, ma2014new, Ma2018, solyaev2019three} and mechanical metamaterials \cite{DellIsola2019} etc.

Solution for the effective shear modulus of the elastic composite with spherical or cylindrical inclusions can be easily obtained in SGET based on the known solution for the Eshelby problem \cite{ma2014new,Ma2018} and homogenization approaches used in the micromechanics of composites. Such type of solutions in SGET are always size-dependent and allow to describe an increase of the effective properties of nanocomposite materials \cite{lurie2011eshelby, ma2014new, Ma2018, solyaev2019three}. Assuming that the matrix is incompressible and that inclusions are rigid we derive then a compact formula that up to the notations represent the size-dependent solution for the effective viscosity of suspensions. 

An approach that is used in the present work can be illustrated by the following classical example. At first, we can consider the known classical solution for  the effective shear modulus $\mu_{clas}^*$ of the elastic particulate composites in the case of small volume fraction of the inclusions, that is given by \cite{christensen2012mechanics}:
\begin{equation}
\label{muclas}
	\mu_{clas}^* = \mu_0 \left( 1 - 
					\frac{15(1-\nu_0)(1-\mu_1/\mu_0)\phi}
					{7 - 5\nu_0 + 2(4 - 5\nu_m)\mu_1/\mu_0} 
				\right)
\end{equation}
where subscripts 0 and 1 denote the properties of matrix and inclusion phases, respecitvely, $\mu_i$, $\nu_i$ are the shear moduli and Poisson ratios of phases, $\phi$ is the volume fraction of inclusions.

Assumption of rigid inclusions and incompressible matrix in the solution (\ref{muclas}) provide us the following result:
\begin{equation}
\label{mu}
	\lim\limits_{\substack{\mu_1 \rightarrow \infty \\ \nu_0 \rightarrow 0.5}}\mu_{clas}^* = \mu = \mu_0 \left( 1 - \frac{5}{2} \phi \right)
\end{equation}

that is the Einstein's formula, in which the analogy between elasticity and hydrodynamics problems allows us to interpret the values of $\mu$ and $\mu_0$ as the dynamic viscosities of suspension and of the base fluid, respectively.

Thus, in the present study we derive the generalized solution of the type (\ref{muclas}) in the framework of the strain gradient elasticity theory and used it then for the evaluation of the effective viscosity of nanofluids using asymptotic procedure of the type (\ref{mu}). We show, that resulting solution predicts the increase of the effective viscosity of suspensions containing smaller particles, that is in line with experimental data for many type of nanofluids \cite{Koca2018, Murshed2017}.  Using differential method\cite{Brinkman1952} we also give a modified solution for the concentrated suspensions and use similar approach to develop the solution for the effective viscosity of suspension with fibrous fillers (e.g. with nanotubes).

The rest part of the paper is organized as follows. In Section \ref{mod} we briefly provide the formulation of the strain gradient elasticity theory and show the relationship that can be established between this theory and corresponding gradient theory of the newtonian viscous fluids (in the case of incompressibility and creeping flow). In Section \ref{eff} we derive the solutions for the effective shear modulus and effective viscosity accounting for size effects. Finally, in Section \ref{res} we provide the comparison between derived solution and known experimental data. 

\section{Model}
\label{mod}
\subsection{Strain gradient elasticity theory}
\label{modA}
In SGET, the strain energy density of the linear elastic material is the function of strain and strain gradients \cite{Mindlin1964}:
\begin{equation}
\label{w}
\begin{aligned}
	W(\pmb{\varepsilon}, \nabla \pmb{\varepsilon}) = 
	\tfrac{1}{2} \,\pmb{\varepsilon} : \textbf{C}:\pmb{\varepsilon} +
	\tfrac{1}{2} \,\nabla 
	\pmb{\varepsilon} \,\vdots \,\textbf{A}\, \vdots\, \nabla \pmb{\varepsilon}
\end{aligned}
\end{equation}
where \textbf{C} is classical fourth-order tensor of elastic constants; \textbf{A} in the sixth-order tensor of gradient elastic moduli (additional material constants of the theory); $\pmb{\varepsilon} = \tfrac{1}{2}(\nabla \textbf{u} + (\nabla \textbf{u})^T)$ is an infinitesimal strain tensor,  $\nabla \pmb{\varepsilon}$ is the strain gradient, $\pmb{u}(\textbf{r})$ is the vector of mechanical displacements at a point $\textbf{r} = \{x_1, x_2, x_3\}$; $\nabla$ is the 3D nabla operator.

Constitutive equations for the classical Cauchy stress tensor $\pmb{\tau}$ (work-conjugate to strain) and third-order double stress tensor $\pmb{\mu}$ (work-conjugate to strain gradients) are given by:
\begin{equation}
\label{CE}
\begin{aligned}
	\pmb{\tau} = \frac{\partial W}{\partial \pmb{\varepsilon}} = 
	\textbf{C}:\pmb{\varepsilon},\quad\qquad
	\pmb{\mu} = \frac{\partial W}{\partial (\nabla \pmb{\varepsilon})} = 
	\textbf{A}\,\vdots\, \nabla \pmb{\varepsilon}
\end{aligned}
\end{equation} 

Equilibrium equations and boundary conditions for the static problems can be derived based on the variational approach in the following form \cite{Mindlin1964}:
\begin{equation}
\label{BVP}
\begin{cases}
	&\nabla \cdot \pmb{\sigma}+ \bar{\pmb{b}}=0, \qquad \textbf{x}\in\Omega\\
	&\textbf{n} \cdot \pmb{\sigma} 
	- \nabla_S \cdot(\textbf{n} \cdot \pmb{\mu}) 
	- H \, \textbf{n} \textbf{n} : \pmb{\mu} = \bar{\pmb{ t}}, 
	\quad or \quad
	\textbf{u} = \bar{\textbf{u}},
	\qquad \textbf{x}\in\partial\Omega\\
	&\textbf{n} \textbf{n} : \pmb{\mu} = \bar{\pmb{ m}}
	\quad or \quad
	\partial_n \textbf{u} = \bar{\pmb{ g}}, \qquad \textbf{x}\in\partial\Omega
\end{cases}
\end{equation} 
where $\pmb{\sigma} =\pmb{\tau} - \nabla \cdot \pmb{\mu}$ is the so-called total stress tensor; $\bar{\pmb{ b}}$ is the body force acted within the volume $\Omega$; $\bar{\pmb{ t}}$, $\bar{\pmb{ m}}$, $\bar{\pmb{ u}}$ and $\pmb{\bar g}$ are the traction, double traction, displacement vector and normal gradient of displacement, respectively, prescribed on the body surface $\partial\Omega$;  \textbf{n} is the vector of the external unit normal to $\partial\Omega$; 
$H = - \nabla_S \cdot \textbf{n}$ is the mean curvature of $\partial\Omega$;  $\nabla_S = \nabla - \textbf{n} \partial_n$ is the surface gradient operator. 

Several form of the constitutive relations (\ref{CE}) can be derived for the second gradient isotropic materials using different assumptions for the structure of the sixth-order tensor of gradient constants $\textbf{A}$ \cite{DellIsola2009, Gusev2015}. In this paper, we will consider the so-called simplified strain gradient elasticity \cite{Askes2011, Askes2009}, for which it is simpler to show the mathematical analogy between statements of the problems for the linear elastic bodies and newtonian viscous liquids (thought, it can be shown, that the use of more general gradient theories of Mindlin's type will lead to the same results). In the considered simplified gradient theory, Cauchy stress tensor and double stress tensor are given by \cite{Askes2011}:
\begin{equation}
\label{CEe}
\begin{aligned}
	\pmb{ \tau}&= \lambda (\nabla\cdot \textbf{u}) \,\textbf{I} 
	+ 2\mu\, \pmb{ \varepsilon}\\
	\pmb{\mu}& = \ell^2 \nabla\pmb{\tau}=\ell^2 \nabla(\lambda 
	(\nabla\cdot \textbf{u}) \,\textbf{I} 
	+ 2\mu\, \pmb{\varepsilon})
\end{aligned}
\end{equation} 
where  $\lambda$, $\mu$ are the classical Lame constants and $\ell$ is the single additional material constant of the theory that can be treated as the intrinsic length scale parameter of material; \textbf{I} is the identity second order tensor.

Note, that the intrinsic length scale $\ell$ is defined by the atomic/molecular structure of media. Identification of such length scale parameters of gradient theories for different elastic materials have been realized up to date using different methods, e.g. based on the comparison between continuum gradient elasticity solutions and first-principles calculations \cite{Stengel2016, maranganti2007novel}, molecular dynamic simulations \cite{gusev2010strain} and experimental data \cite{liebold2016comparison}. If $\ell=0$, then the presented formulation (\ref{w})-(\ref{CEe}) will have the classical from of the theory of elasticity without length scale parameters.

Substituting (\ref{CEi}) into (\ref{BVP}.1) we obtain the equillibrium equations in terms of displacements in the following form \cite{Askes2011}:
\begin{equation}
\label{GE}
\begin{aligned}
	(1-l^2 \nabla^2) \big((\lambda+\mu) \nabla \nabla \cdot \textbf{u} 
	+ \mu \Delta \textbf{u}\big) + \bar{\pmb{b}}= 0
\end{aligned}
\end{equation}

One can see, that in SGET the equilibrium equations (\ref{GE})  are of the fourth order and there exist an extended number of boundary conditions (see (\ref{BVP})). In particular, in the problems with inhomogeneous structures there arise the additional continuity conditions prescribed with respect to the normal gradient of displacements $\partial_n \pmb{u}$ (see \cite{ma2014new, solyaev2019three}). These conditions results in the so-called effects of "boundary layers", that arise at the interfaces, where the strain in the softer phase becomes as small as in stiffer phase. As result, the overall stiffness of the inhomogeneous material increases. Moreover, as the amount of boundaries in the composite materials with smaller inclusions become higher, the positive size effects for the effective properties of nanocomposite materials can be described using SGET\cite{lurie2011eshelby, ma2014new, Ma2018, solyaev2019three}. 
In the present paper, we propose to transfer these well known results in solid mechanics to the hydrodynamics problems with incompressible viscous fluids containing  small size inclusions. 

\subsection{Relationship to the gradient theory of viscous fluid}
\label{modB}
The quasi-static gradient theory of incompressible viscous fluids can be obtained based on the presented formulation of SGET using the following procedure (note, that this procedure is similar to those one used in classical theories \cite{Batchelor1974, christensen2012mechanics}). At first, we can assume that the mass continuity equation remain classical even in the case of non-classical form of strain energy density (\ref{w}). Thus, in the case of incompressibility we can use the requirement: $\nabla\cdot \textbf{u} = 0$. Additionally, we should take into account that the Poisson's ratio in the case of incompressibility tends to 0.5, i.e. that the bulk modulus $K = \lambda+2\mu/3$ and the first Lame parameter $\lambda$ tend to infinity. In this case, in (\ref{CEe}) and in (\ref{GE}) terms $(\lambda \cdot\nabla \textbf{u})$ are indeterminate and can be written in terms of the reactive hydrostatic pressure $p$. As result, the constitutive equations (\ref{CEe}) are reduced to:
\begin{equation}
\label{CEi}
\begin{aligned}
	\pmb{ \tau}&= -p \,\textbf{I} 
	+ 2\mu\, \pmb{ \varepsilon}\\
	\pmb{\mu}& = \ell^2(- \nabla p \,\textbf{I} 
	+ 2\mu \,\nabla\pmb{\varepsilon})
\end{aligned}
\end{equation} 

and governing equations (\ref{GE}) become:
\begin{equation}
\label{GEi}
\begin{aligned}
	(1-l^2 \nabla^2) (\mu \nabla^2 \textbf{u} - \nabla p) + \bar{\pmb{b}}= 0
\end{aligned}
\end{equation}
 
Next, assuming the standard analogy for the notations in the solid and fluid mechanics \cite{christensen2012mechanics, Batchelor1974}, we can use the replacement rule in (\ref{BVP}), (\ref{CEi}), (\ref{GEi}), according to which the displacements $\textbf{u}$ should be replaced by velocity, strain $\pmb{\varepsilon}$ and strain gradient $\nabla\pmb{\varepsilon}$ become the strain rate and strain rate gradient, respectively, stresses $\pmb{\tau}$ and $\pmb{\mu}$ should be treated as the classical and high-order viscous stresses, the shear modulus $\mu$ should be replaced by the dynamic viscosity. Additional high-order boundary conditions in (\ref{BVP}) can be defined as the so-called generalized adherence conditions\cite{Fried2006}. The analog for the strain energy (\ref{w}) in fluid mechanics will be the rate of  energy dissipation due to viscosity, which will depends in this case on the strain rate and its gradients. 

As result, we will obtain the form of equations (\ref{BVP}), (\ref{CEi}), (\ref{GEi}) that is similar to those one derived in the second gradient fluid mechanics \cite{Fried2006}. The single difference is that the so-called "vectorial hyperpressure" \cite{Fried2006} is expressed here through the pressure gradient, that is the consequence of the made constitutive assumption in (\ref{CEe}).

Based on the described mathematical analogy we can use the known solutions developed in SGET for the analysis of fluid mechanics problems accounting for size effects, as the intrinsic length scale parameter $\ell$ will remain in such solutions (this parameter measures the strength of the fluid’s adherence to the boundaries in second gradient fluid mechanics\cite{Fried2006}). Namely, we consider the solutions for the effective shear modulus of composite materials with spherical and cylindrical inclusions. Using procedure described above we can replace in these solutions the elasticity material constants on their hydrodynamics analogs and use it to describe the effective viscosity of suspensions accounting for the fillers size. Such solutions are given in the next section.

\section{Size-dependent effective shear modulus and effective viscosity}
\label{eff}
Based on the direct homogenization approach, tensor of the effective elastic moduli $\textbf{C}^*$ of the composite material with dilute concentration of the inclusions can be obtained based on the following relation \cite{aboudi2013}:
\begin{equation}
\label{Cx}
\begin{aligned}
	\textbf{C}^* = \textbf{C}_0 
	+ \phi (\textbf{C}_1 - \textbf{C}_0) \textbf{T},
\end{aligned}
\end{equation}

where $\phi$ is volume fraction of inclusions, subscripts 0 and 1 define the properties of the matrix and inclusions, respectively, $\textbf{T}$ is the fourth-order strain concentration tensor, which for the ellipsoidal inclusion can be represented through the properties of phases and Eshelby tensor $\textbf{S}$ as follows \cite{aboudi2013}:
\begin{equation}
\label{T}
\begin{aligned}
	\textbf{T} = \left(\textbf{I} 
	+ \textbf{S} \,\textbf{C}_0^{-1} (\textbf{C}_1 - \textbf{C}_0)\right)^{-1}
\end{aligned}
\end{equation}

In turn, the Eshelby tensor $\textbf{S}$ for spherical inclusions in SGET have been evaluated recently in \cite{Ma2018, ma2014new}. Thought the components of Eshelby tensor are position dependent in SGET (strain and stress fields are not uniform inside the inclusion even in the case of uniform external loading  \cite{lurie2011eshelby, ma2014new, Ma2018, solyaev2019three}), for the small volume fractions of inclusions it is appropriate to use the averaged values of these components over the inclusion volume, like it was proposed by Ma and Gao \cite{ma2014new} and validated in our recent work \cite{lurie2018comparison} using energy based homogenization approach and numerical simulations. 

Thus, from Refs. \cite{Ma2018, ma2014new} we take the strain gradient elasticity solution for the averaged components of Eshelby tensor for spherical inclusion:
\begin{equation}
\label{S}
\begin{aligned}
	S_{ijkl} =  s_1 \delta_{ij}\delta_{kl} 
	+ &s_2(\delta_{ik}\delta_{jl} + \delta_{il}\delta_{jk} )\\[5pt]
	s_ 1 =  \frac{5\nu_0-1}{15(1-\nu_0)}&
	\Phi(\bar d),\quad
	s_ 2 =  \frac{4-5\nu_0}{15(1-\nu_0)} 
	\Phi(\bar d)\\[5pt]
	\Phi(\bar d) = 1
	+ \frac{3}{\bar d^3} 
	&\left(4-\bar d^{\,2} - (2+\bar d)^2 e^{-\bar d}\right)
\end{aligned}
\end{equation}

and for cylindrical inclusion (non-zero copmonents):
\begin{equation}
\label{Sc}
\begin{aligned}
	&S_{\alpha\beta\gamma\delta} =  
	c_1 \delta_{\alpha\beta}\delta_{\gamma\delta} 
	+ c_2(\delta_{\alpha\gamma}\delta_{\beta\delta} 
	+ \delta_{\alpha\delta}\delta_{\beta\gamma} ),\quad
	S_{\alpha3\beta3} =  c_3\delta_{\alpha\beta},\quad
	S_{\alpha\beta33} =  c_4\delta_{\alpha\beta}, \\[5pt] 
	c_ 1 &=  \frac{4\nu_0-1}{8(1-\nu_0)}
	\Gamma(\bar d),\quad
	c_ 2 =  \frac{3-4\nu_0}{8(1-\nu_0)} 
	\Gamma(\bar d),\quad
	c_3 = \frac{1}{4}\Gamma(\bar d),\quad
	c_4 = \frac{\nu_0}{2(1-\nu_0)} \Gamma(\bar d),\\[5pt]
	&\Gamma(\bar d) = 1 - 2 K_1
	(\tfrac{\bar d}{2}) I_1(\tfrac{\bar d}{2})
\end{aligned}
\end{equation}

where the latin indexes $i,j,k,l$ take values 1, 2, 3, and greek indexes takes  values 1 and 2; $\nu_0$ is the Poisson's ratio of matrix; $\delta_{ij}$ is Kroneker symbol; $K_1$ and $I_1$ are the modified Bessel functions of the first and second kinds of the first order, respectively; $\bar d = d/\ell$ is the relative diameter of inclusion normalized to the length scale parameter of the matrix $\ell$.

Using (\ref{S}), (\ref{T}) and (\ref{Cx}) we can derive the following dilute approximation for the effective shear modulus of particulate composite:
\begin{equation}
\label{mux}
\begin{aligned}
	\mu_s^* &= \mu_0 + \frac{\phi (\mu_1 - \mu_0) \mu_0} 
	{\mu_0 + 2 s_2 (\mu_1 - \mu_0)}
\end{aligned}
\end{equation}

Based on the asymptotic procedure similar to (\ref{mu}) and taking into account the definitions of $s_2$ (\ref{S}), we derive then the solution for the composite materials with incompressible matrix and rigid spherical inclusions:
\begin{equation}
\label{muxl}
\begin{aligned}
	\lim\limits_{\substack{\mu_1 \rightarrow \infty \\ 
	\nu_0 \rightarrow 0.5}}\mu_s^* 
	= \mu_s (\mu_0, \phi, \bar d) = \mu_0\left(1 
	+ \frac{5} {2\Phi(\bar d)}\phi\right)
\end{aligned}
\end{equation}

Accounting for the mentioned in Section \ref{modB} relationship between gradient models of elastic solids and viscous fluids, we can use (\ref{muxl}) to evaluate the effective dynamic viscosity $\mu_s$ of dilute suspensions with spherical particles. Note, that in the case of relatively large particles ($\bar d \rightarrow \infty$) it is valid that $\lim\limits_{\bar d\rightarrow \infty}\Phi(\bar d) =1$ and solution (\ref{muxl}) reduces to the classical Einstein formula (\ref{mu}). Thus, we define solution (\ref{muxl}) as the generalized Einstein's formula that takes into account the influence of the particles size.

Solution for the concentrated suspensions can be easily obtained based on the approach proposed in fluid mechanics\cite{Brinkman1952}, which is also known in the composite mechanics as the differential homogenization scheme \cite{aboudi2013}. Thus, we can evaluate an increase of the effective viscosity $d\mu$ due to addition of small amount of particles $d\phi$ as follows:
\begin{equation}
\label{dmu}
\begin{aligned}
	\mu+ d\mu = \mu_s (\mu, d\phi', \bar d), \quad 
	d\phi'=\frac{d\phi}{1-\phi}
\end{aligned}
\end{equation}
where $d\phi'$ is actual volume fraction of added inclusions that takes into account the corresponding decrease of the base fluid volume.

From (\ref{dmu}) and (\ref{muxl}) we obtain the following differential equation with the initial condition, which implies that without fillers the effective viscosity equals to those one of base fluid:
\begin{equation}
\label{mupde}
\begin{aligned}
	\begin{cases}
		\frac{d\mu}{d\phi} = \frac{5 \mu} {2(1-\phi)\Phi(\bar d)}\\
		\mu|_{\phi=0} = \mu_0
	\end{cases}
\end{aligned}
\end{equation}

Solution of (\ref{mupde}) provide us the following assessment for the effective viscosity of concentrated suspensions of spherical particles:
\begin{equation}
\label{mudif}
\begin{aligned}
	\mu_{s,c} = \frac{\mu_0}{(1 - \phi)^{\frac{5}{2\Phi(\bar d)}}}
\end{aligned}
\end{equation}

This solution (\ref{mudif}) also takes into account the effect of inclusion size. Note, that in the case of large particles ($\bar d \rightarrow \infty$) this solution reduces to the famous Brinkman's solution for the concentrated suspensions\cite{Brinkman1952}.
 
For the composites with dilute concentration of cylindrical inclusions we can use (\ref{Cx}), (\ref{T}) and (\ref{Sc}) to derive the following solutions for the effective longitudinal and transverse shear moduli:
\begin{equation}
\label{mux1223}
\begin{aligned}
	\mu_L^* &= \mu_0 + \frac{\phi (\mu_1 - \mu_0) \mu_0} 
	{\mu_0 + c_3 (\mu_1 - \mu_0)}, \qquad
	\mu_T^* &= \mu_0 + \frac{\phi (\mu_1 - \mu_0) \mu_0} 
	{\mu_0 + c_2 (\mu_1 - \mu_0)}
\end{aligned}
\end{equation}

Based on (\ref{mux1223}) we can find the asymptotic solution of the type (\ref{mu}) for the effective viscosity of fibrous suspensions. However, here we should take into account that for any cylindrical inclusion there exist two planes of the longitudinal shear and single plane of the transverse shear, thus for the media containing randomly oriented fibers we can derive the following approximation:
\begin{equation}
\label{muxc}
\begin{aligned}
	\lim\limits_{\substack{\mu_1 \rightarrow \infty \\ 
	\nu_0 \rightarrow 0.5}}\frac{2\mu_L^* + \mu_T^*}{3} 
	= \mu_c (\mu_0, \phi, \bar d) = \mu_0\left(1 
	+ \frac{4} {\Gamma(\bar d)}\phi\right)
\end{aligned}
\end{equation}

For the concentrated fibrous suspensions one can use the similar differential method given by (\ref{dmu}), (\ref{mupde}) and obtain the following solution:
\begin{equation}
\label{muxcc}
\begin{aligned}
	\mu_{c,c} = \frac{\mu_0}{(1-\phi)^\frac{4}{\Gamma(\bar d)}}
\end{aligned}
\end{equation}

Note, that formulas (\ref{muxc}), (\ref{muxcc}) reduces to the classical analogs in the case of large size inclusions, because it is valid that $\lim\limits_{\bar d\rightarrow \infty}\Gamma(\bar d) =1$.


\section{Results and discussion}
\label{res}
\begin{figure}
\textbf{a}\includegraphics[width=0.45\textwidth]{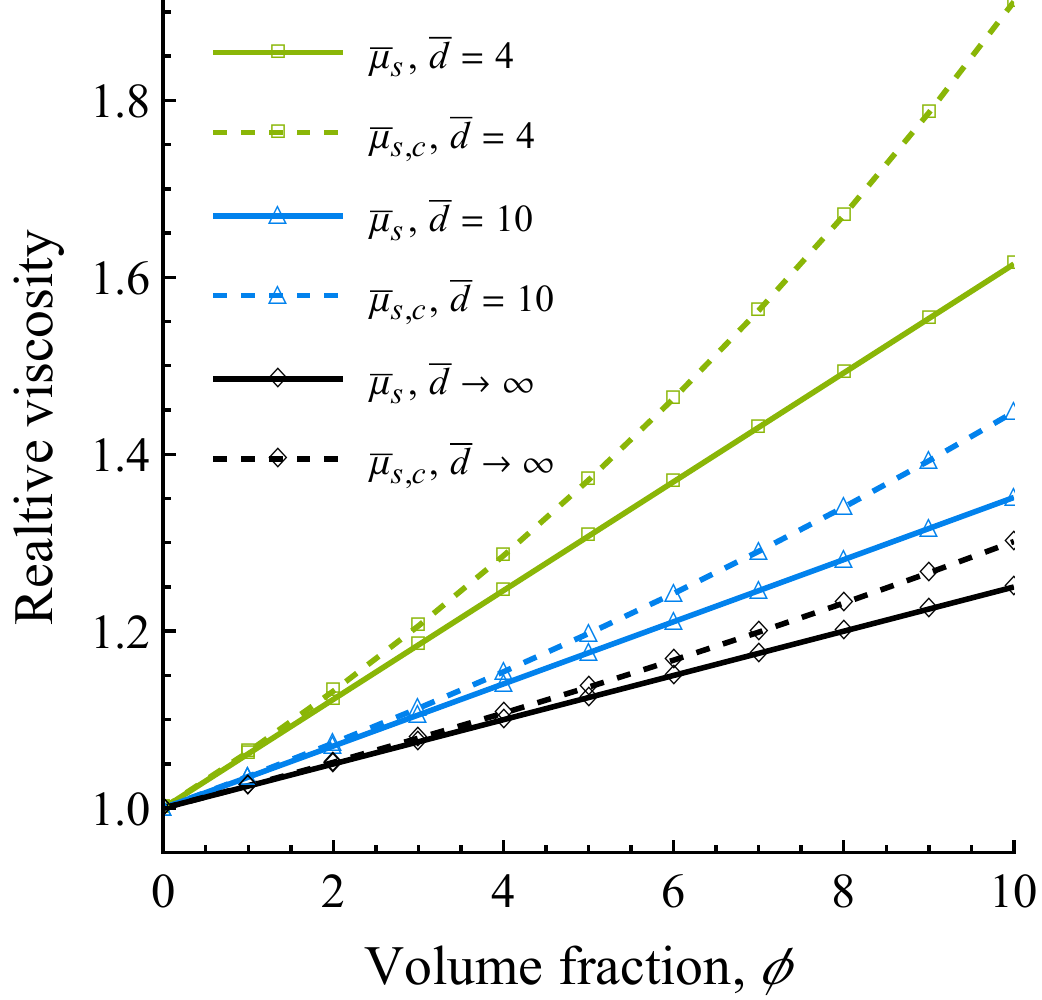}\qquad
\textbf{b}\includegraphics[width=0.45\textwidth]{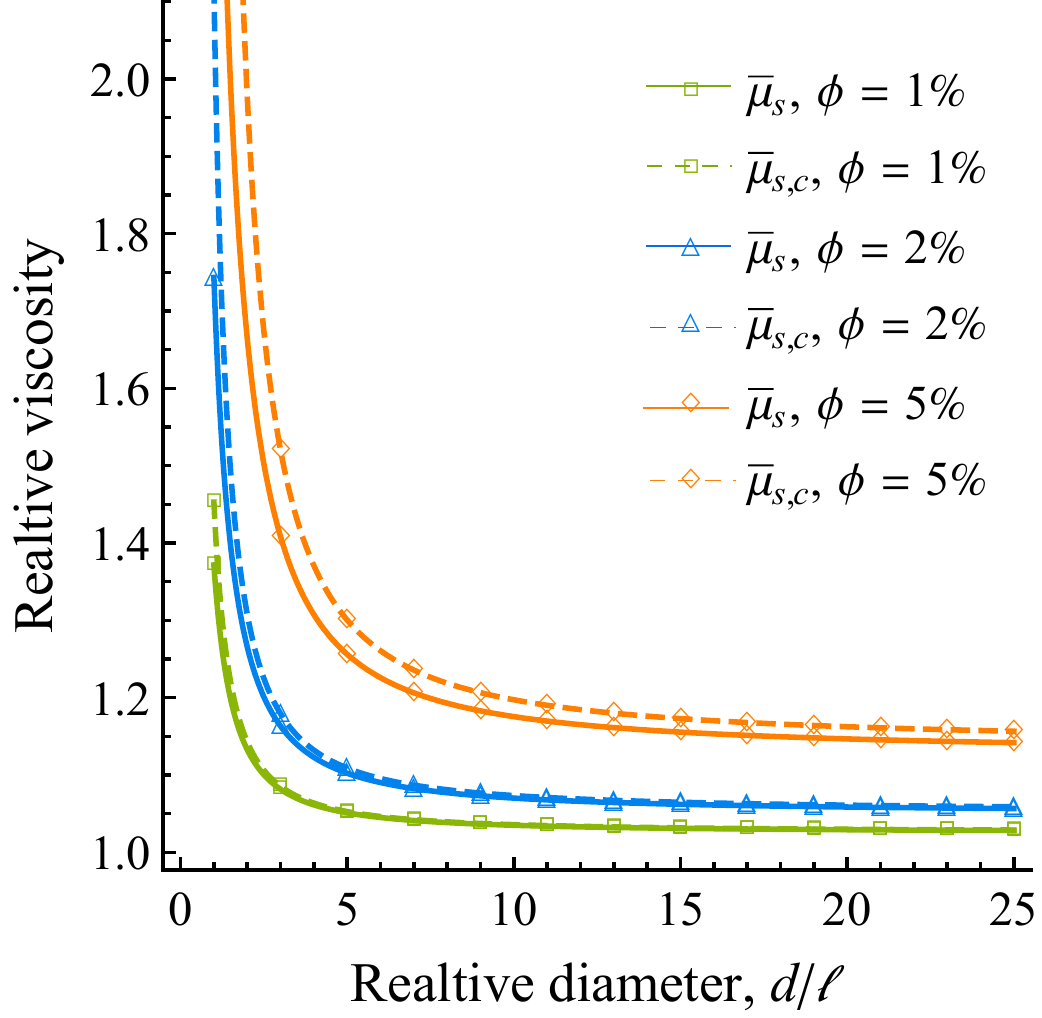}
\caption{\label{fig1}Dependence of relative effective viscosity of suspension on the volume fraction (a) and diameter (b) of spherical particles. Solid lines -- generalized Einstein solution (\ref{muxl}) for the dilute suspensions, dashed lines --  generalized Brinkman solution (\ref{mudif}) for concentrated suspensions}%
\end{figure}

Obtained solutions for the effective viscosity (\ref{muxl}) and (\ref{mudif}) are illustrated in Fig. \ref{fig1}. Here we show the dependence of relative effective viscosity of the dilute ($\bar\mu_{s}=\mu_{s}/\mu_0$) and concentrated ($\bar\mu_{s,c} = \mu_{s,c} /\mu_0$) suspensions on the volume fraction and relative diameters of particles. In the case of large inclusions ($\bar d\rightarrow \infty$) presented solutions are reduced to the classical Einstein's and Brinkman's solutions\cite{einst1096, Brinkman1952}, that are shown by the black solid and black dashed  lines in Fig. \ref{fig1}a, respectively. For the finite values of ratio $d/\ell$, obtained generalized solutions predict a positive size effects -- the viscosity of suspensions with smaller particles becomes higher. Influence of the inclusions size becomes significant in the case of rather small inclusions, which  diameter is less then $d\approx10\ell$ (i.e. $\bar d <10$, see Fig. \ref{fig1}b). The length scale parameter $\ell$ is the additional material constant of the model and we do not have an \textit{apriori} assessment for its values, however, we will evaluate it below based on the comparison of the modeling results with known experimental data. Note, that solutions for the fibrous suspensions $\mu_{c}$ (\ref{muxc}) and $\mu_{c,c}$ (\ref{muxcc}) have a very similar form as shown in Fig. 1 and we do not present it here. 

\begin{figure}
\textbf{a}\includegraphics[width=0.45\textwidth]{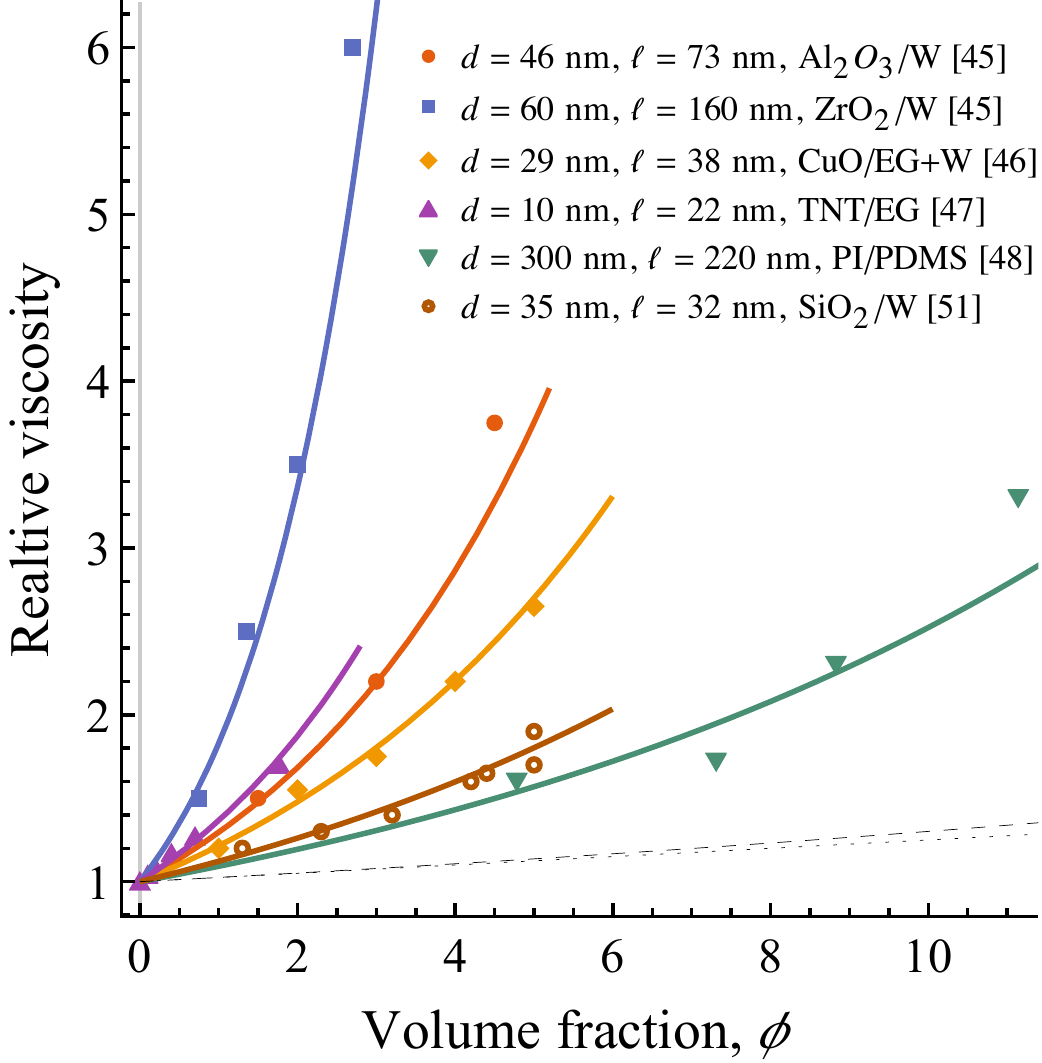}\qquad
\textbf{b}\includegraphics[width=0.45\textwidth]{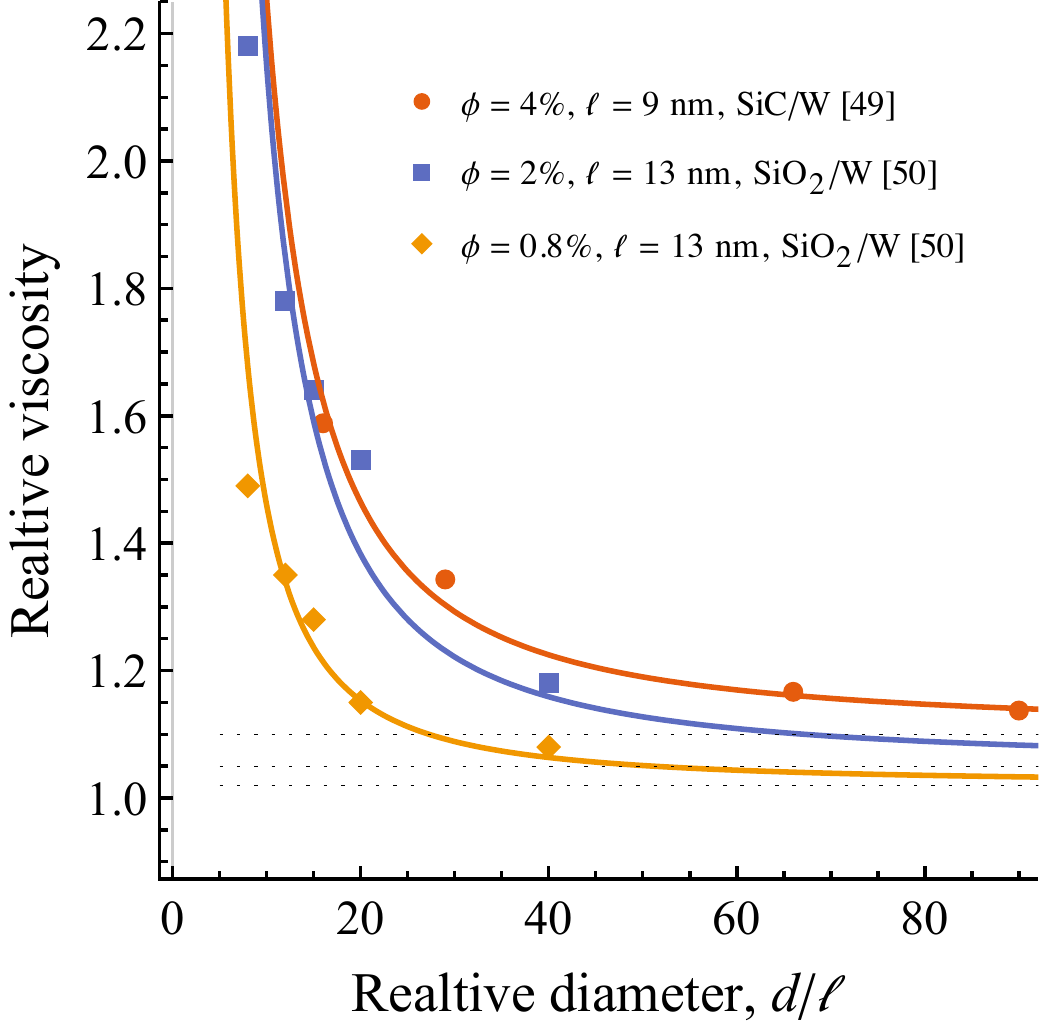}
\caption{\label{fig2}Comparison between fitted modeling results (lines) and experimental data (points) for the effective viscosity of different nanofluids. Dependences of relative effective viscosity on the volume fraction of nanoparticles and nanotubes (a) and on the relative diameter of particles (b) are presented. Models of the concentrated suspensions (\ref{mudif}),  (\ref{muxcc}) are used in calculations. Classcial Einstein and Brinkman solutions are shown by black dotted and dashed lines, respectively}%
\end{figure}

\begin{figure}[]
\textbf{a}\includegraphics[width=0.45\textwidth]{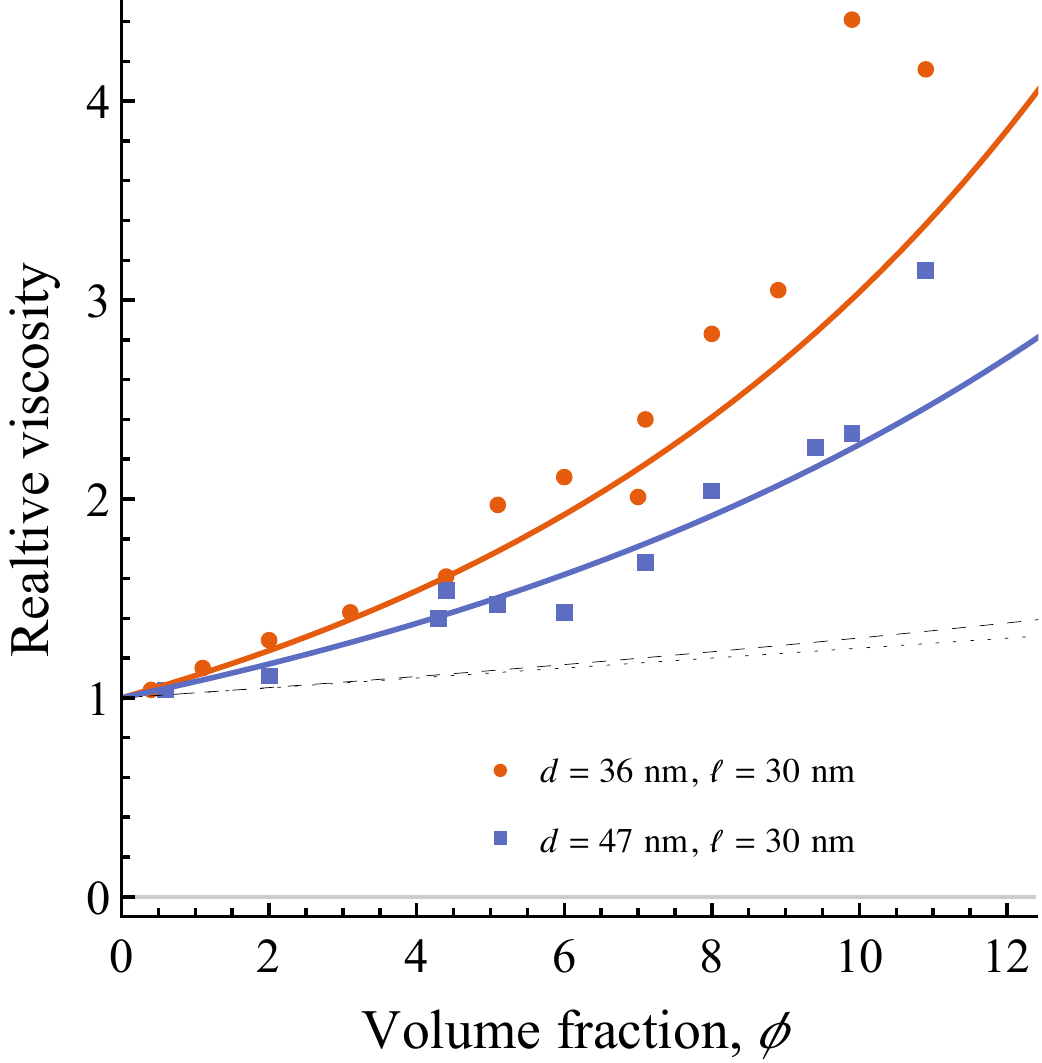}\qquad
\textbf{b}\includegraphics[width=0.45\textwidth]{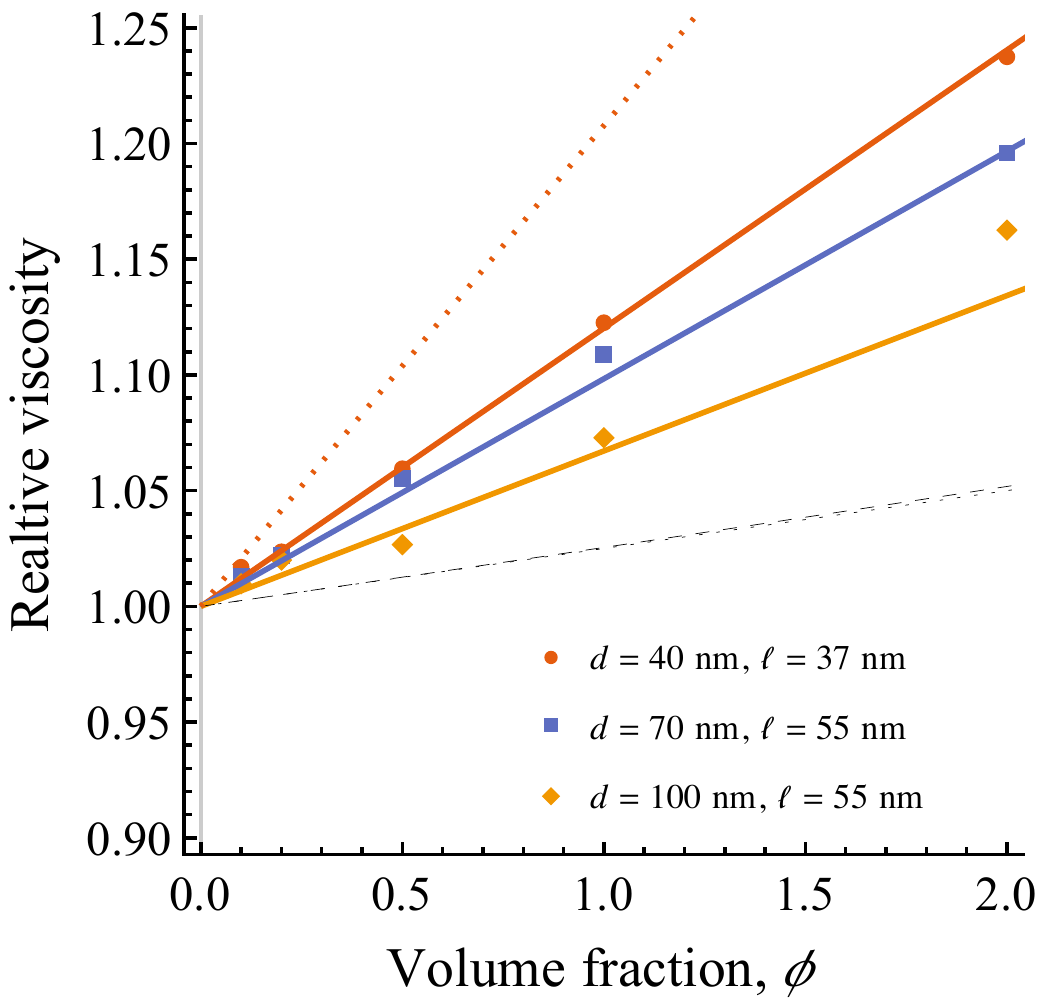}
\caption{\label{fig3}Comparison between fitted modeling results (lines) and experimental data (points) for the effective viscosity of Al$_2$O$_3$/water (a) and Fe/water (b) nanofluids\cite{nguyen2008, esfe2014} . Dependence of relative effective viscosity on the volume fraction of nanoparticles is presented. Models of the concentrated (\ref{mudif}) and dilute (\ref{muxl}) suspensions are used. Classical Einstein and Brinkman solutions are shown by black dotted and dashed lines, respectively}%
\end{figure}

Comparison of the developed models with experiments is shown in Figs. \ref{fig2} and \ref{fig3}. Experimental values of relative viscosity of different nanofluids were taken from Refs.\cite{nguyen2008, williams2008, namburu2007, chen2009, semenov2020, timofeeva2010,jia2009, chevalier2007, esfe2014}, in which the base fluids were the water (W), etylenglycol (EG), \textit{polydimethylsiloxane} (PDMS) and the fillers were the different oxide ceramics, Fe and polyimide nanoparticles and titanate nanotubes (TNT). In Fig. \ref{fig2}a we show the dependence of the effective viscosity of different nanofluids on the particles and nanotubes volume fraction. For the calculations we used here the model of the concentrated suspensions with spherical particles $\mu_{s,c}$ (\ref{mudif}) and with fibers $\mu_{c,c}$ (\ref{muxcc}). These models were fitted to the experimental data by using single additional parameter -- length scale parameter $\ell$, while the values of volume fraction, particles/nanotubes diameter and the viscosity of base fluid were known. In such a way we identify the length scale parameter $\ell$ for the considered suspensions (shown in the figures). It is seen, that found values of $\ell$ for the considered nanofluids can be less or more than the diameter of inclusions, however usually they have the same order. The highest values of this parameter realize in the case of very fast increase of the effective viscosity in the dilute suspensions (see e.g. the result for the ZrO$_2$/water suspension in Fig. \ref{fig2}a). Solutions for the dilute concentration (\ref{muxl}), (\ref{muxc}) can be also fitted to the presented data, but in the smaller range of volume fractions. Classical Einstein's and Brinkman's solutions (thin black lines in Fig. \ref{fig2}a) cannot describe the presented experimental results.

In Fig. \ref{fig2}b we show the dependence of the effective viscosity on the particles size. One can see rather good agreement between experiments and  fitted model (\ref{mudif}). In this case we also used the unknown model parameter $\ell$ to describe the experimental values of the effective viscosity that is much higher than predicted by the classical solutions (black dashed lines in Fig. \ref{fig2}b). It is interesting to note, that for the SiO$_2$/water suspensions\cite{jia2009} with different volume fraction of particles we identified the same length scale parameter $\ell = 13$ nm. This means, that in this case the model allows us to describe two effects: the dependence of the effective viscosity on the volume fraction and on the diameter of particles. 

Another example for the description of the suspension with the same composition but with different volume fraction and size of particles is shown in Fig. \ref{fig3}. To describe the viscosity of the Al$_2$O$_3$/water nanofluid\cite{nguyen2008} we use the generalized Brinkman's solution (\ref{mudif}) and single value of the length scale parameter $\ell=30$ nm, which allows us to appoximate the experiments with volume fraction of inclusions up to $\phi=5$ \% and with inclusions diameters $d = $ 36 nm and 47 nm (Fig. \ref{fig3}a). For the suspension with larger particles this value of $\ell$ is also valid for the volume fraction up to $\phi = 10$ \%, however, for the smaller particles the difference between modeling and experimental data becomes higher with increase of volume fraction $\phi>5$ \%. Thus, it may be needed to involve more complex models to describe such suspensions with high volume fraction of small particles \cite{lurie2018comparison}. 

In Fig. \ref{fig3}b we use the experimental data for the nanofluids Fe/water\cite{esfe2014}. Here we used the generalized Einstein's solution (\ref{muxl}), as the volume fraction of inclusions were not more then 2\%. It was found that the single length scale parameter $\ell = 55$ nm allows to describe the experiments for the larger inclusion ($d = $ 70 nm and 100 nm), however, to describe the suspension with smallest inclusions ($d = $ 40 nm) we have to reduce it to $\ell = 37$ nm. In other case model overestimates the effective viscosity (see red dotted line in Fig. \ref{fig3}b). Therefore, we can assume that the interactions (adherence) between liquid and smallest particles in the suspensions were changed due to some physical or chemical reasons, e.g. due to oxidation of particles surfaces. Detailed experimental studies of such effects are needed. However, even in such case the presented model can be useful -- it can be used to  check the possible change of adhesion between base liquid and fillers of different size based on the measured values of the effective viscosity of suspension.

Note, that presented in Figs. \ref{fig2} and \ref{fig3} experimental values for the increased relative viscosity of nanofluids is typical \cite{Koca2018, Munyalo2018}. Such results cannot be described by the classical models: nor Einstein and Brinkman classical solutions nor by the other known models that do not take into account the influence of the particles diameter. Of course, there exist different reasons of size effects in nanofluids, including agglomeration\cite{
krishna2015modeling,Ganguly2009} and Brownian motion\cite{Masoumi2009}, however, it seems that the presented second gradient models are also valid and the described effects can make a significant impact on the properties of the certain type of nanofluids. These effects are similar to those one, that are usually related to the particles capping layers \cite{Hosseini2010}, however, here we propose a continuum description of such effects without introduction of any additional phases in the suspensions.

It is also important that developed solutions have clear physical background as it bases on the homogenization approaches for the inclusion problems in the framework of the second gradient continuum theory. Since the initial elasticity solutions for the composites (\ref{mux}), (\ref{mux1223}) are derived assuming that the strain energy density of phases depends on the strain and strain gradients, the developed solution for the effective viscosity takes into account that the rate of energy dissipation in fluids depends on the first and second gradients of velocity. Such formulation of the fluid mechanics model can be justified for the small scale flows\cite{Fried2006}, and namely for the flows around a very small particles in nanofluids, where the high velocity gradients arise. 

The direct reason of the size effects that are predicted by the obtained solutions lie in the more general boundary (continuity) conditions that are used in the gradient theories and that was called the "generalized adherence conditions" in second gradient fluid mechanics\cite{Fried2006}. Detailed discussion about such type of boundary conditions and related boundary effects that arise in gradient theories of solids and fluids can be found elsewhere \cite{Fried2006, fried2008continuum, lurie2011eshelby, ma2014new, Ma2018, solyaev2019three}. Here, it is important to note, that presented solutions allow to take into account the size effects assuming that there exist some volume of fluid adhered to the particles boundaries. In the case of nanoparticles, the total volume of this adhered liquid becomes very high and make a significant impact on the effective viscosity of nanofluid. Such adhesion-type effects depends on the both contacted materials -- liquid and solid, such that the length scale parameter $\ell$ should be treated as the measure of interactions between them. Moreover, the value of $\ell$ may depend on the temperature and other physical and chemical effects. Experimental measurement and theoretical prediction of the length scale parameters for different fluids is the field for the further research. It seems, that the value of $\ell$ can depend on the equivalent molecular diameter of the base fluid, like it was proposed in the empirical model by Corcione\cite{Corcione2010}. Molecular dynamic simulations can be also useful for the evaluation of the length scale parameters for different suspensions, because the strength of adherence can be directly studied in such approaches that were also useful in gradient theories of elastic solids\cite{gusev2010strain}. 

Note, that presented solutions are derived assuming the quasi-static creeping flow, thus the effects of strain rate are not captured and these solutions are valid in the case of small strain rates. However, the rate effects can be also  taken into account considering the full statement of the second gradient fluid mechanics and numerical simulations.

\section{Conclusions}
\label{con}
In this paper we derived the generalized Einstein's and Brinkman's solutions for the effective viscosity of the dilute and concentrated suspensions of solid spheres and fibers. Solutions are derived based on the mathematical analogy between second gradient theories for the linear elastic solids and newtonian liquids. Initial elasticity solution is developed as the dilute approximation for the effective shear moduli of the composites and used then for the evaluation of effective viscosity assuming incompressible behavior of matrix and rigid behavior of solid inclusions. 

It is shown that presented solutions allow to describe the effect of increase of the effective viscosity of suspensions with small particles, that was widely observed in the experiments with nanofluids. Additional length scale parameter of the model is identified in this work based on the known experimental data for different nanofluids.



%
%

%


\bibliography{refs.bib}

\end{document}